\documentclass[mathleft
]{an}
\usepackage{graphicx}
\usepackage{times}
\newcommand\dnu{\Delta\nu}
\newcommand\dnumoy{\langle\Delta\nu\rangle}
\newcommand\deltanunu{\Delta\nu(\nu)}
\newcommand\seconde{\Delta_2\nu}
\newcommand\dnup{\Delta\nu\ind{even}}\newcommand\dnui{\Delta\nu\ind{odd}}
\newcommand\dnucomb{\Delta\nu_{{\tiny{\vert\vert\vert}}}(\nu)}
\newcommand\dnuavoid{\Delta\nu\ind{avoid}}
\newcommand\ordre{n'}
\newcommand\ordredeux{\alpha}
\newcommand\fcomb{{\mathcal{F}_{\Delta\nu}}}

\newcommand\numax{\nu\ind{max}}

\newcommand\filtre{\delta\nu\ind{F}}

\newcommand{\ind}[1]{_{\mathrm{#1}}}

\def\muHz{\,$\mu$Hz}

\def\m2s2{\,m$^{2}$\,s$^{-2}$} 

\def\aap{A\&A}
\newcommand\Kepler{\emph{Kepler}}

\begin{document}

\Pagespan{9999}{}
\Yearpublication{2010}%
\Yearsubmission{2010}%
\Month{11}%
\Volume{000}%
\Issue{000}%

\title{An alternative to mode fitting}

\author{B. Mosser\inst{1}\fnmsep\thanks{Corresponding author:
  \email{benoit.mosser@obspm.fr}\newline}
}
\titlerunning{An alternative to mode fitting}
\authorrunning{B. Mosser}
\institute{
LESIA, CNRS, Universit\'e Pierre et Marie Curie, Universit\'e Denis Diderot, Observatoire de Paris, 92195 Meudon, France
}

\received{March, 2010}
\accepted{  }
\publonline{later}

\keywords{stars: interiors -- stars: evolution -- stars: oscillations -- stars: individual, \object{HD 49385}, \object{HD 49933}, \object{HD 181420}, \object{Procyon} -- techniques: photometry}

\abstract{%
The space mission CoRoT provides us with a large amount of high-duty cycle long-duration observations. Mode fitting has proven to be efficient for the complete and detailed analysis of the oscillation pattern, but remains time consuming. Furthermore, the photometric background due to granulation severely complicates the analysis. Therefore, we attempt to provide an alternative to mode fitting, for the determination of large separations. With the envelope autocorrelation function and a dedicated filter, it is possible to measure the variation of the large separation independently for the ridges with even and odd degrees.
The method appears to be as accurate as the mode fitting. It can be very easily implemented and is very rapid.
}

\maketitle

\section{Introduction}

A new era in asteroseismology has been opened by the mission CoRoT, that provides us with high-quality photometric time series. The Fourier spectra have revealed a large number of solar-like oscillations in main-sequence stars and subgiants, plus oscillations in thousands of giants. In comparison to previous ground-based observations, observation duration is long, with a duty cycle of the order of 90\,\%. The ground-based network dedicated to Procyon (\cite{2008ApJ...687.1180A}) has reached a similar duty cycle, but only for the central 10\,days of the network observation.

Both the analysis of these solar-like oscillations and the scientific output crucially depend on the quality of the data. Automated pipelines  based on different methods have been setup (\cite{2010A&A...511A..46M}, \cite{2009A&A...506..465H}, \cite{2009CoAst.160...74H}). In order to extract precise eigenfrequencies, modes fitting is required, as made e.g. in \cite{2009A&A...507L..13B}. The performance of this time-consuming method appears to be limited by the background level. Therefore, we investigate an alternative approach. We focus our analysis on the measurement of the large separation and propose to extend the method presented in  \cite{2009A&A...508..877M}, hereafter MA09.

\cite{2006MNRAS.369.1491R} have presented how to analyze
solar-like oscillations with the autocorrelation envelope function
(EACF). This function searches directly for the signature of the
large separation in the autocorrelation of the time series,
without any prior on the oscillation excess power. The EACF
measures directly the acoustic radius of the star. The delay of
correlation being shorter than the mode lifetimes, the correlation
is able to stack coherent signal in the time series. For rapidity,
the autocorrelation of the time series is performed as the Fourier
spectrum of its Fourier spectrum. For efficiency and detailed
analysis, a filter helps to select a given frequency range in the
Fourier spectrum. MA09 have analyzed the performance of the
detection of the large separation with the EACF. \cite{ma10} have
presented the structure of the automated pipeline based on the
EACF.

The method has proven to be efficient at low signal-to-noise ratio (\cite{2009A&A...506...33M}).
A positive detection, with the maximum amplitude of the EACF above the threshold level, gives at least the measurement of the mean value of the large separation. At larger signal-to-noise ratio (SNR), the method provides the variation of the large separation. For high-quality data, it gives also a test for identifying the degrees of the modes.

Deriving the variation of the large separation with frequency is highly interesting for stellar modeling  (\cite{2010A&A...514A..31D}). Contrary to eigenfrequencies, the large separations are much less sensitive to the influence of the upper atmospheric layers. However, as shown by many studies (e.g. \cite{1993A&A...274..595P}), the large separation variation $\deltanunu$ may depend on the degree of the mode. A method for deriving $\deltanunu$ for individual degree is highly desirable. We will consider the measurement of $\dnup$ and $\dnui$, corresponding to the even and odd values of the degree. \cite{2010CoAst.161....3B} have proposed to make use of the frequencies of the ridge centroids in order to define $\dnup$ and $\dnui$. In this paper, we propose a more direct approach, based on the EACF but with a dedicated filtering.

The method is presented in Section~\ref{method}. We present its coherence: it first gives the mean value $\dnumoy$ of the large separation, then the variation $\deltanunu$, then the values $\dnup$ and $\dnui$ obtained for the even and odd ridges separately.
Different results are presented and discussed in Sect.~\ref{analysis}. We show that we easily reproduce results based on the time consuming mode fitting.
Section~\ref{conclusion} is devoted to conclusions.


\section{Method\label{method}}

We intend to perform an analysis of the even and odd ridges
separately. Therefore, the basic concept of the method consists in
introducing a comb structure in the filter of the EACF. We present
the recipe used for extracting such an information without a
priori. We use the EACF to first  define the comb and then derive
the measurement of $\dnup$ and $\dnui$. We do not base the
analysis on the solar case, since the Sun does not present
significantly different $\dnup$ and $\dnui$. Instead, we consider
the case of HD\,49385, a G-type subgiant observed with CoRoT that
shows significantly different even and odd ridges
(\cite{2010A&A...515A..87D}): the large separations between radial
modes are about 56\muHz, whereas for $\ell=1$ modes they can be as
low as 50\muHz.

\subsection{Analysis of the comb structure}

\begin{figure}
\centering
\includegraphics[width=7.98cm]{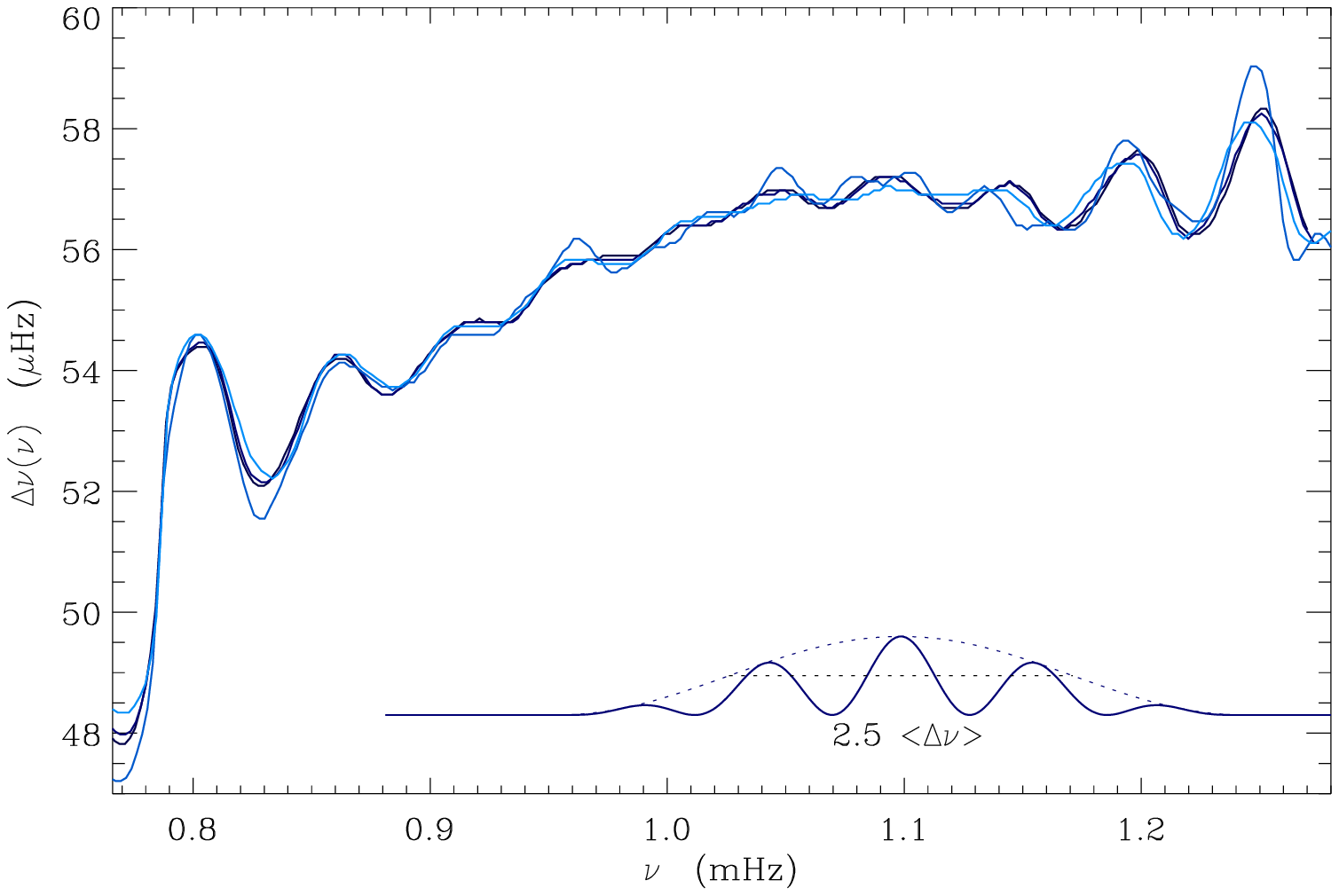}
\caption{$\dnucomb$ for HD\,49385 based on 5 different large separations varying from 54 to 58\muHz. The inset shows the filter $\fcomb$ for $\dnumoy = 56$\muHz\ and $\filtre = 2.5\,\dnumoy$.
\label{fig_fonctionridgeHD49385}}
\end{figure}

MA09 have determined that the EACF with a filter width equal to the full-width at half-maximum (FWHM) of the excess power envelope centered on the frequency $\numax$ of maximum amplitude gives the mean value $\dnumoy$ of the large separation. With a narrower filter, with a typical FWHM of $2\dnumoy$, one can extract the variation of the large separation with frequency $\deltanunu$. In order to analyse independently the even and odd ridges, we introduce a comb structure in the filter $\fcomb$:

\begin{equation}
\!\!\!\!\left\lbrace
\begin{array}{rlc}
 \fcomb (\nu')  &= \displaystyle{1\over 4} & \left[1  +  \cos \pi\displaystyle{\nu'\over \filtre} \right]
 \left[1  +  \cos 2\pi\displaystyle{\nu'\over \dnumoy} \right] \\
 & & \mathrm{for\ } |\nu'| \le \filtre \\
 & & \\
 \fcomb (\nu')  &= 0 &   \mathrm{for\ } |\nu'| > \filtre \\
\end{array}
\right.
\end{equation}
where $\nu'$ is the frequency shift to a central frequency $\nu$
and $\filtre$ is the FWHM of the Hanning filter. We call
$\dnucomb$ the function $\deltanunu$ obtained from the
autocorrelation with such a filter
(Fig.~\ref{fig_fonctionridgeHD49385}).

The definition of $\dnucomb$ has the disadvantage to introduce the mean value $\dnumoy$ of the large separation into the analysis. Subsequently, it may be suspected to bias the result. We have shown that it is not the case, by comparing the values of $\dnucomb$ obtained with different comb filters based on different large separations (Fig.~\ref{fig_fonctionridgeHD49385}). We have superimposed the results for 5 different values of the comb period: $\dnumoy$, $\dnumoy\pm1$\muHz\ and  $\dnumoy\pm2$\muHz.
The $\pm2$\muHz\ shift represents a relative variation of about 4\,\%, much greater than the precision of the measurement of $\dnumoy$. Since the different values of $\dnucomb$, even when obtained with strongly biased filters, are in fact very similar, we conclude that taking a filter based on the mean value of the large separation is adequate.

\begin{figure}
\centering
\includegraphics[width=7.98cm]{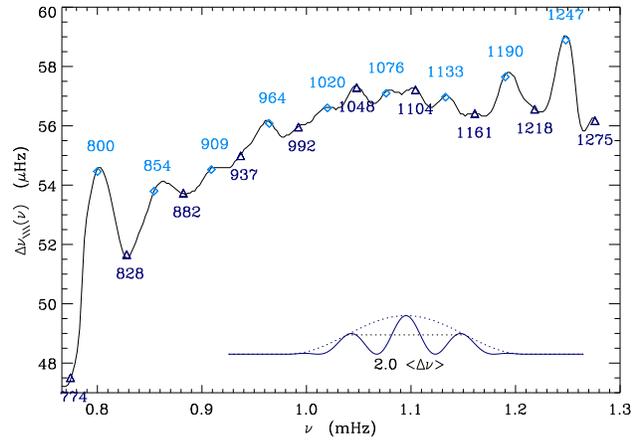}
\caption{Same as Fig. \ref{fig_fonctionridgeHD49385}, with the superimposition of the estimates of the eigenfrequencies and large separations obtained with Eq. \ref{tassoul_like}. The filter is based on the mean large separation $\dnumoy$; its FWHM is changed compared to Fig.~\ref{fig_fonctionridgeHD49385} in order to show its reduced influence.
\label{fig_peakridgeHD49385}}
\end{figure}

\begin{figure}
\centering
\includegraphics[width=7.98cm]{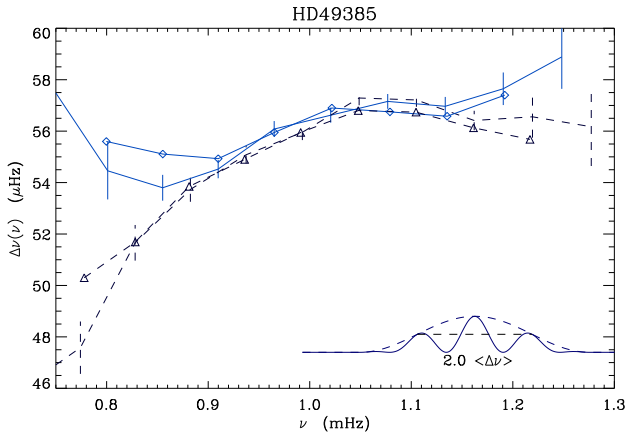}
\caption{$\dnup$ and $\dnui$ for HD\,49385. The solid lines (respectively dashed lines) correspond to the even (reps. odd) ridge; 1-$\sigma$ error bars are given with the same linestyles. We have superimposed the values presented in \cite{2010A&A...515A..87D} derived from the mode fitting: diamonds correspond to $\ell=0$, triangles to $\ell=1$. Error bars of the fitted values, not shown, are equivalent to the error bars derived from the EACF.
\label{fig_autodeltanuridgeHD49385}}
\end{figure}

\subsection{Ridges}

The function $\dnucomb$ previously defined is continuous. In order to derive $\dnup$ and $\dnui$, we have to identify the even and odd ridges and to locate the eigenfrequencies independently of any mode fitting. The ridge identification was already performed in MA09 or by \cite{2010CoAst.161....3B}, so that we are left with the estimate of the eigenfrequencies.


In order to perform this, we simply use the EACF to measure the first order variation of the large separation:
\begin{equation}
\dnu \simeq \dnumoy +\ordredeux\   \ordre
\label{1erordre}
\end{equation}
with $n'=n-n_0+\ell/2$ and $n_0$ the radial order associated with the nearest radial eigenfrequency to $\numax$. From Eq.~(\ref{1erordre}), we can derive proxies of the $\ell=0$ and $\ell=1$ eigenfrequencies:
\begin{equation}
\nu_{n,0} \simeq \nu_0 + \dnup \ \ordre + \ordredeux_0 \  {\ordre^2\over 2}
\label{tassoul_like}
\end{equation}
\begin{equation}
\nu_{n,1} \simeq \nu_1 + \dnui \ \ordre + \ordredeux_1 \  {\ordre^2\over 2}
\label{tassoul_like_i}
\end{equation}
All parameters are direct outputs of the EACF analysis, except the
constant terms $\nu_0$ and $\nu_1$. Their precise values, close to
$\numax$, are derived from the minimization of the residuals
between the observed and estimated eigenfrequencies.

Figure~\ref{fig_peakridgeHD49385} shows the values of the large separation $\dnucomb$ estimated at the frequencies determined by Eq. (\ref{tassoul_like}). We note that $\dnup$ and $\dnui$ correspond to local extrema of the function $\dnucomb$. This justifies a posteriori that a precise location of the eigenfrequencies is not necessary to derive $\dnup$ and $\dnui$. We can then compare $\dnup$ and $\dnui$ to their fitted values, namely the values derived by the mode fitting.

\subsection{Performance}

The performance of the method, expressed by the error bars on $\dnup$ and $\dnui$, depends basically on the scaling of the noise (Eqs. 3-6 of MA09). Compared to the results presented in MA09, the error bars are essentially multiplied by a factor 2 since the comb pattern divides by 2 the efficient number of points in the filter.

Tests with different filters have shown that rectangular or triangular shapes induce too many discontinuities in the function $\dnucomb$. We therefore prefer to consider Hanning filters, as in MA09, with a FWHM varying from 2 to 4 $\dnumoy$. Smaller widths give the finest resolution, but require a high signal-to-noise ratio, indicated by high values of the EACF.

\begin{figure}
\centering
\includegraphics[width=7.98cm]{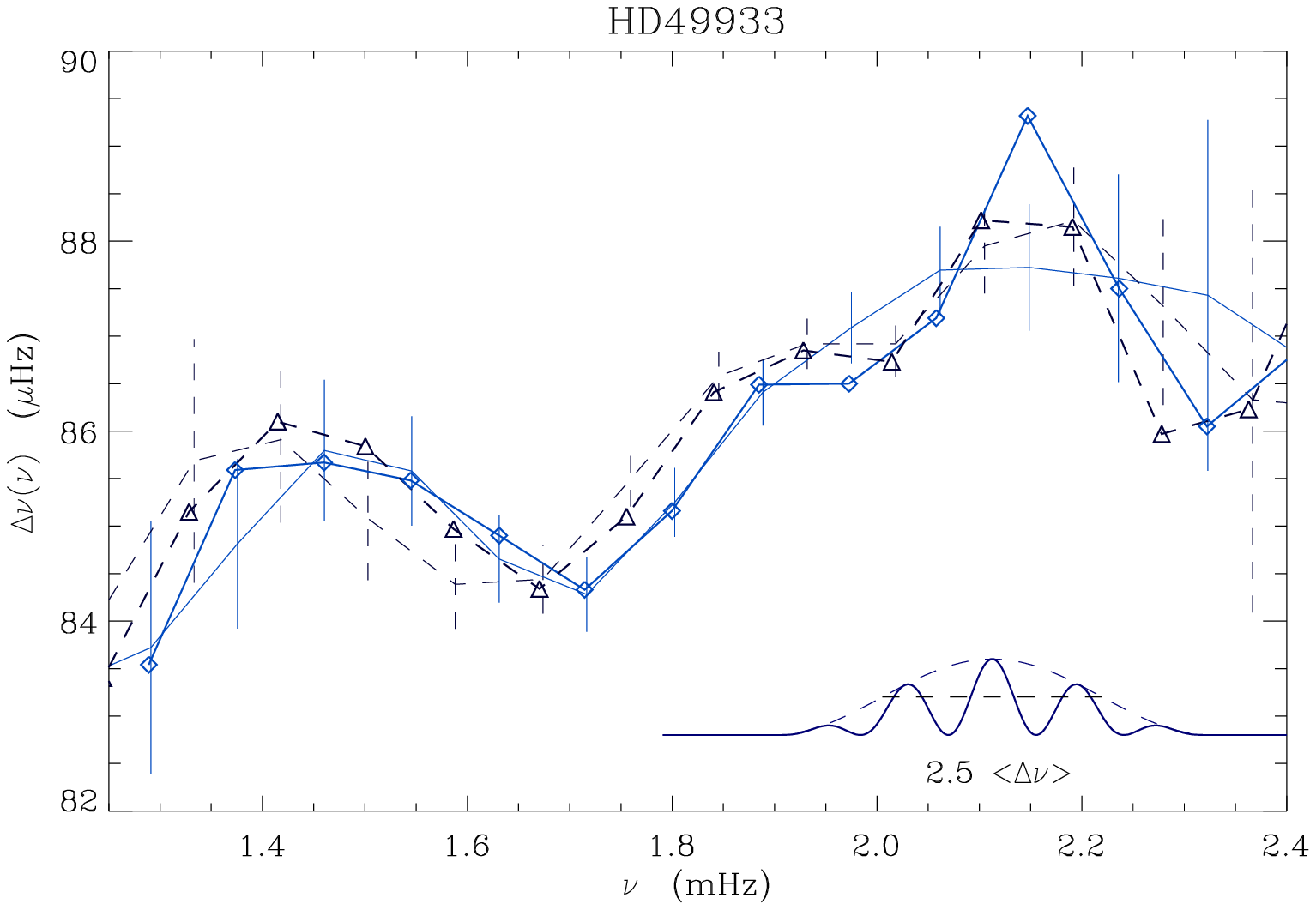}
\caption{\label{fig_autodeltanuridgeHD49933}
Same figure as Fig.~\ref{fig_autodeltanuridgeHD49385}, comparing $\dnup$ and $\dnui$ for HD\,49933 with fitted values determined by  \cite{2009A&A...507L..13B}.}
\end{figure}

\begin{figure}
\centering
\includegraphics[width=7.98cm]{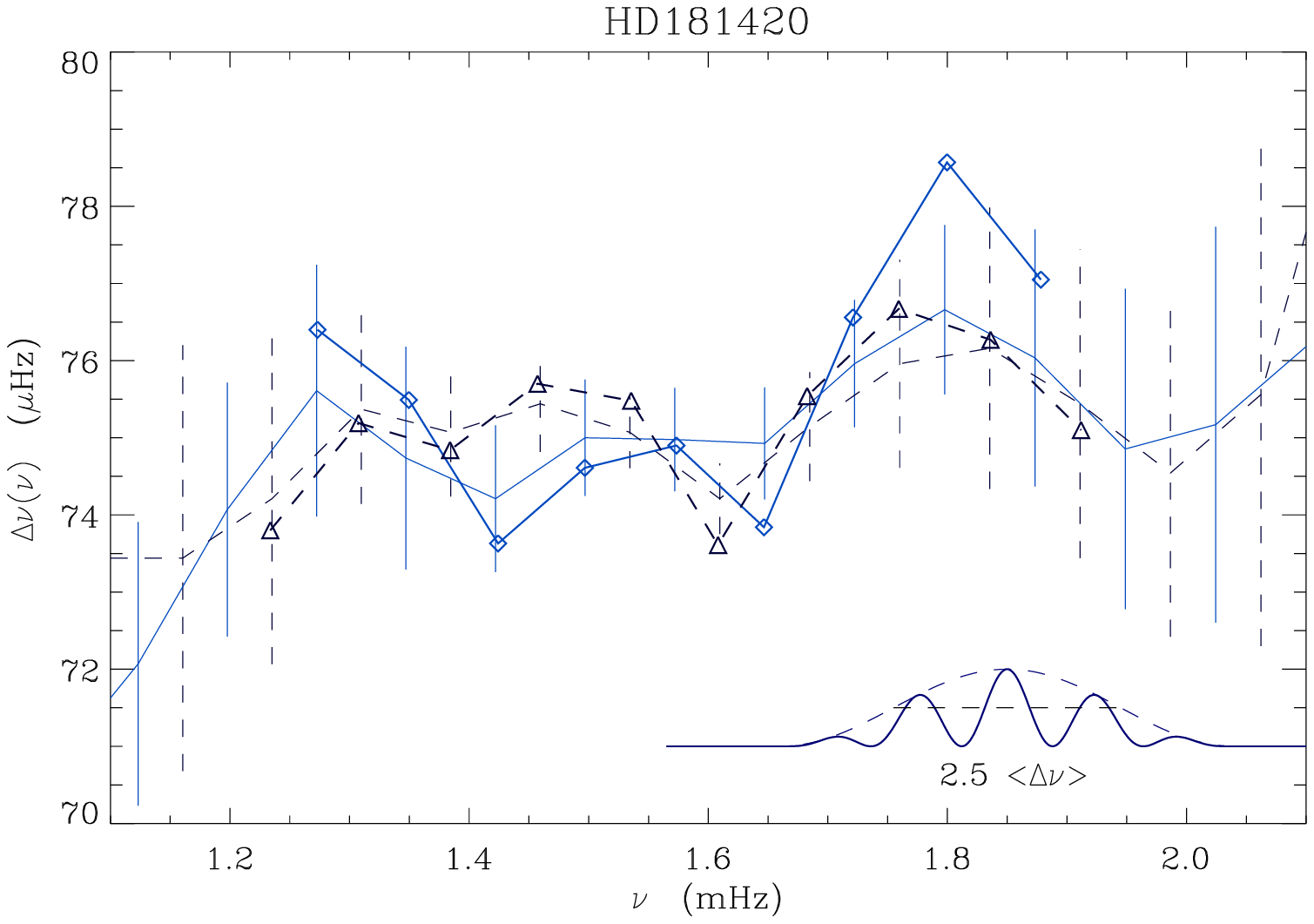}
\caption{\label{fig_autodeltanuridgeHD181420}
Same figure as Fig.~\ref{fig_autodeltanuridgeHD49385}, comparing $\dnup$ and $\dnui$ for HD\,181420 with fitted values determined by \cite{2009A&A...506...51B}. Data are more noisy, but the agreement is as clear as for oscillation spectra with higher signal-to-noise ratio.}
\end{figure}

\begin{figure}
\centering
\includegraphics[width=7.98cm]{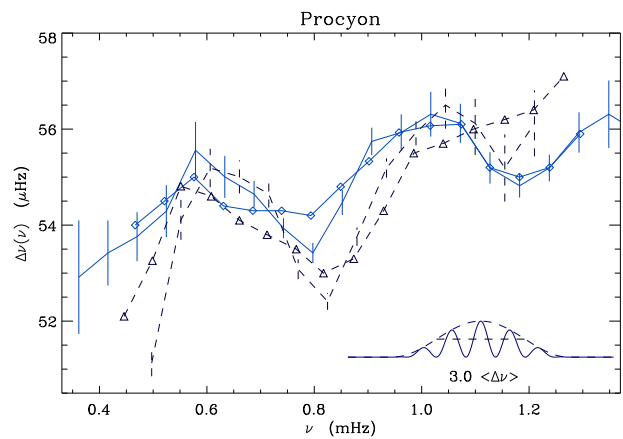}
\caption{\label{fig_autodeltanuridgeProcyon}
Same figure as Fig.~\ref{fig_autodeltanuridgeHD49385}, comparing $\dnup$ and $\dnui$ for Procyon with the centroids of the ridges (Fig. 11a of \cite{2010ApJ...713..935B}).
A larger filter has been used to compare our results to the smoothed determination provided by these centroids.}
\end{figure}

\begin{figure}
\centering
\includegraphics[width=7.98cm]{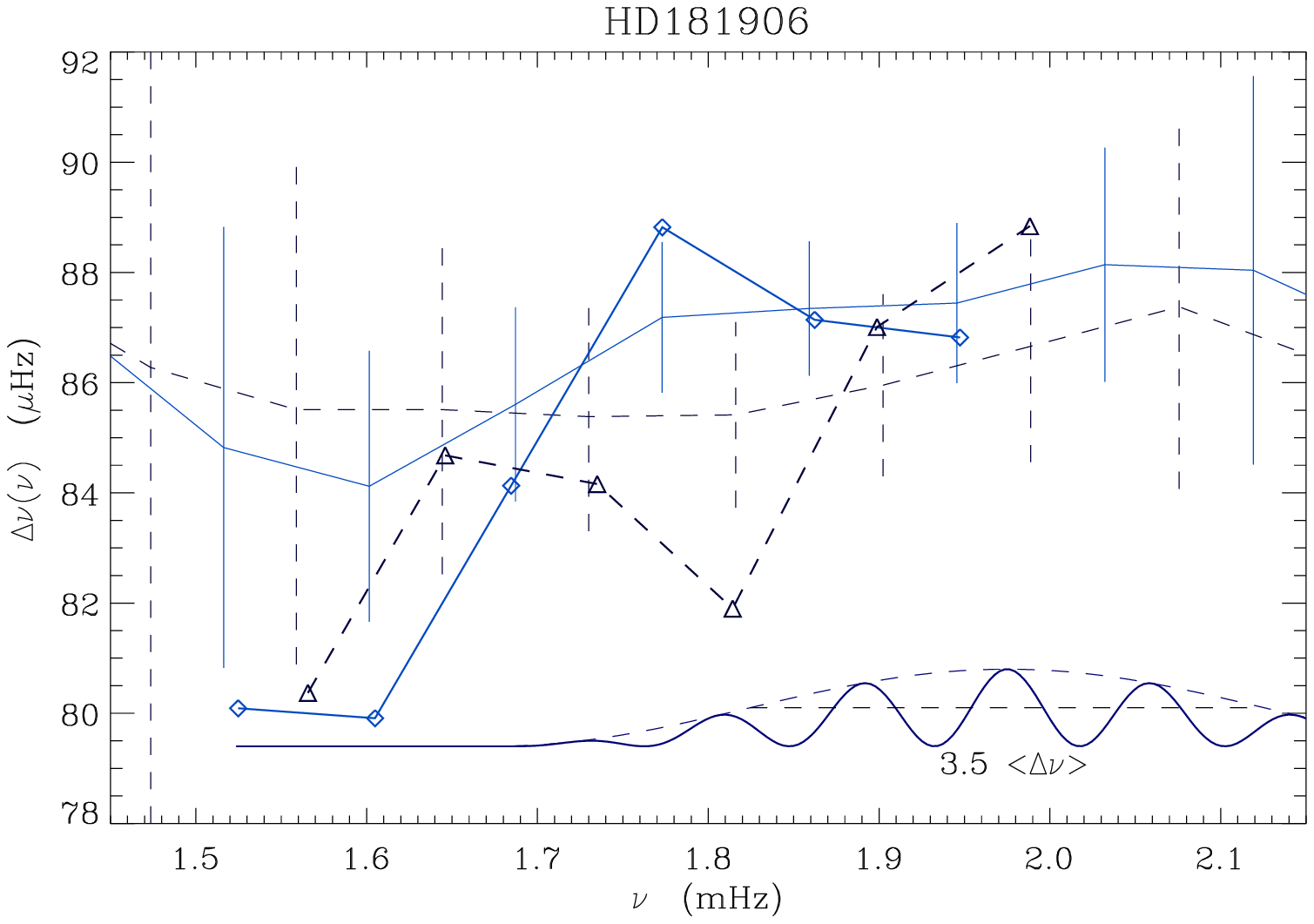}
\caption{\label{fig_autodeltanuridgeHD181906}
Same figure as Fig.~\ref{fig_autodeltanuridgeHD49385}, comparing $\dnup$ and $\dnui$ for HD\,181906 with fitted values determined by \cite{2009A&A...506...41G}. }
\end{figure}

\section{Discussion\label{analysis}}

\subsection{A few results}

We have compared for different stars  $\dnup$ and $\dnui$ obtained with this method or derived from the mode fitting. The agreement lies within the error bars (Fig.\,\ref{fig_autodeltanuridgeHD49385} to \ref{fig_autodeltanuridgeProcyon}). We note that the error bars obtained with the EACF are equivalent to the 1-$\sigma$ error bars of the fitted eigenfrequencies. For clarity, we do not reproduce them in the figure.

For HD\,49385 (Fig.~\ref{fig_autodeltanuridgeHD49385}), the EACF analysis is clearly able to disentangle the very different values of $\dnup$ and $\dnui$ due to avoided crossings (\cite{2010A&A...515A..87D}). For HD\,49933, the method exhibits both the significant modulation of $\deltanunu$  and the difference between $\dnup$ and $\dnui$.

\subsection{At low signal-to-ratio}

For HD\,181420, observed with a lower signal-to-noise ratio, the agreement remains remarkable (Fig.~\ref{fig_autodeltanuridgeHD181420}). The values of $\dnup$ and $ \dnui$ extracted from the EACF have smoother variation than given by the mode fitting. Due to the value of error bars, one cannot decide if the difference is due to the smoothing provided by the filter of the EACF or to spurious variation in the fitted values. On the other hand, the EACF provides values for $\dnup$ and $\dnui$ in a larger frequency range.
By the way, we confirm the identification of the degree of the ridge proposed by MA09, corresponding to scenario 1 of \cite{2009A&A...506...51B}.

It is possible to use a larger filter in order to analyze low SNR light curves. We have performed the analysis of $\dnucomb$ for the CoRoT target HD 181906 (\cite{2009A&A...506...41G}). The difference between $\dnup$ and $\dnui$ (Fig. \ref{fig_autodeltanuridgeHD181906}) is clearly shown, but the method does not provide a way to identify the ridges.
We believe that the clean measurement of $\dnup$ and $\dnui$ may be useful for comparison with a forthcoming modeling of this star.


\begin{figure}
\centering
\includegraphics[width=7.98cm]{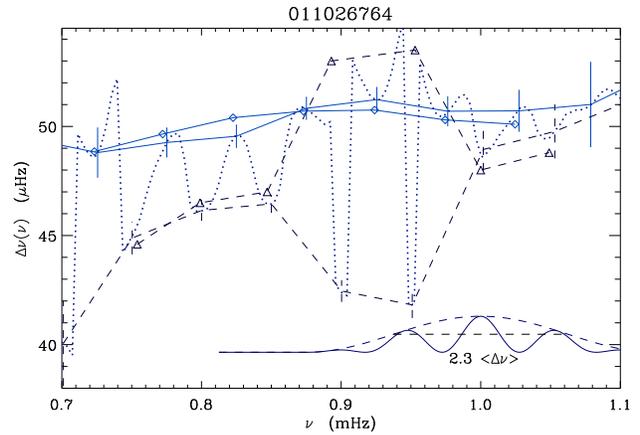}
\caption{\label{fig_autodeltanuridge011026764}
Same figure as Fig.~\ref{fig_autodeltanuridgeHD49385}, comparing $\dnup$ and $\dnui$ for the star KIC\,011026764 observed with \Kepler, with fitted values determined by \cite{2010ApJ...713L.169C}. The function $\dnucomb$ has been superimposed (dotted lines), with discontinuities in the frequency range where the regularity of the oscillation pattern of $\ell=1$ modes is broken.}
\end{figure}

\begin{figure}
\centering
\includegraphics[width=7.98cm]{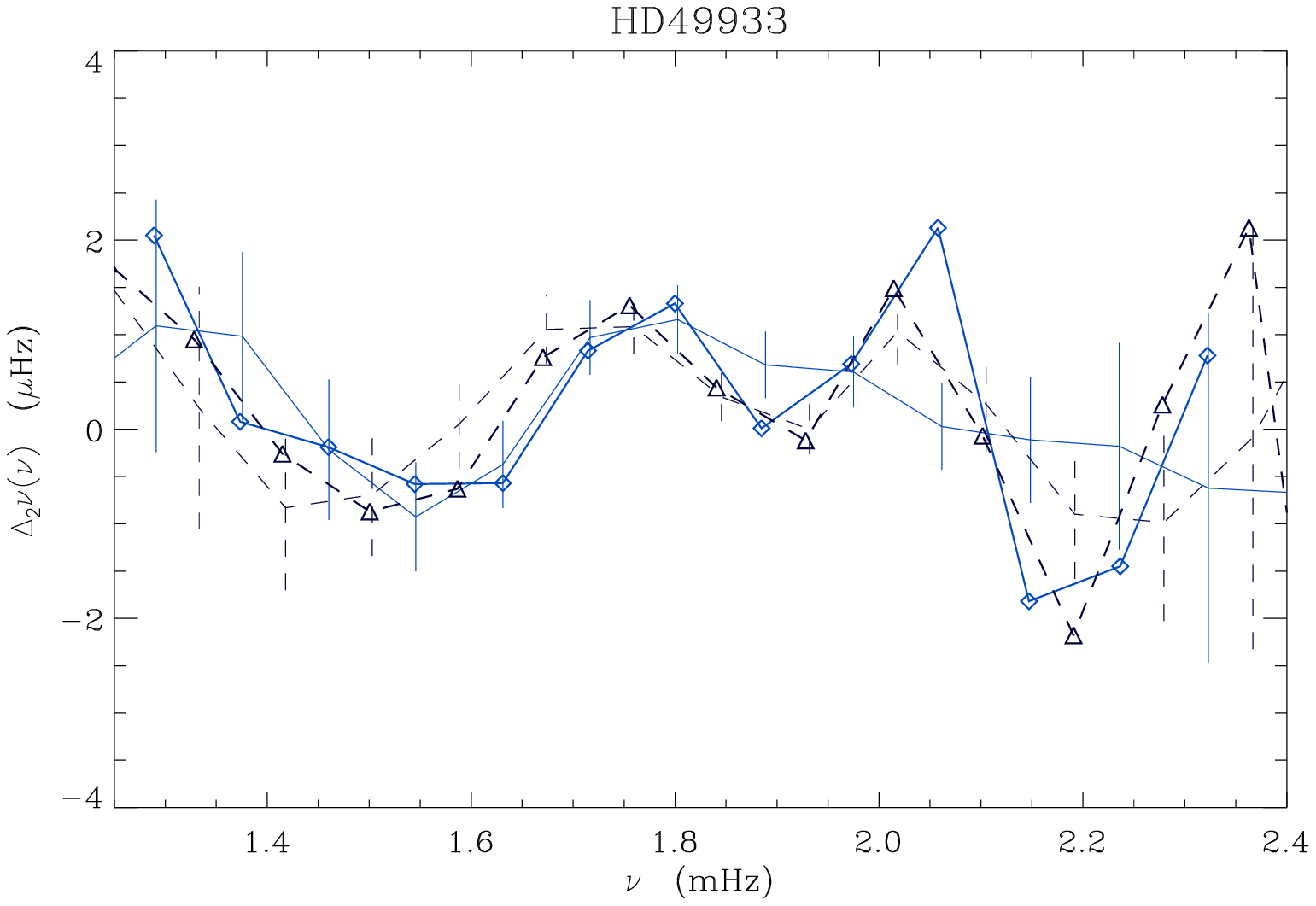}
\caption{\label{fig_autodeltanu2_HD49933}
$\seconde\ind{even}$ and $\seconde\ind{odd}$ for HD\,49933, presented as $\dnu$ in Fig.~\ref{fig_autodeltanuridgeHD49385} and compared to the fitted second differences derived from \cite{2009A&A...507L..13B}}
\end{figure}

\begin{figure}
\centering
\includegraphics[width=7.98cm]{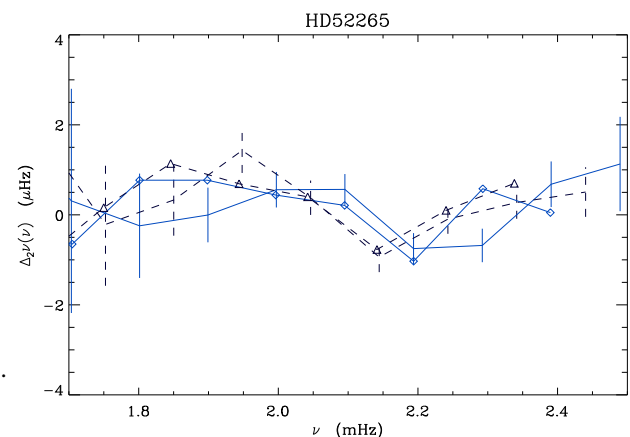}
\caption{\label{fig_autodeltanu2_52265}
Same as Fig. \ref{fig_autodeltanu2_HD49933} for HD\,52265, with comparison to the results from \cite{Ballot2010}.
}
\end{figure}

\subsection{Avoided crossings}

The presence of avoided crossings may complicate the measurement
of $\dnucomb$, since they perturb the regular Tassoul-like
pattern. We have tested this effect on the evolved star
KIC\,11026764 observed by \Kepler\ and analyzed by
\cite{2010ApJ...713L.169C}. This star shows mixed modes, with a
complex pattern in the frequency range [0.88, 0.95\,mHz]. We show
that the method is sensitive to the irregularity of the
oscillation pattern due to avoided crossings. The discontinuities
in $\dnucomb$ indicate clearly the presence of the mixed modes
(Fig.~\ref{fig_autodeltanuridge011026764}). Instead of pointing
the perturbed value $\dnuavoid$ of the large separation at the
avoided-crossing frequency, the method gives $2\dnup-\dnuavoid$.

\subsection{Second difference}

Measuring the large separations gives access to further important asteroseismic typical frequencies.
The so-called second differences
\begin{equation}
\!\!\!\seconde_{n,\ell} = \nu_{n+1, \ell} - 2 \, \nu_{n,\ell} + \nu_{n-1, \ell} = \dnu_{n+1,\ell} - \dnu_{n,\ell}
\label{seconde}
\end{equation}
can be estimated directly. These second differences are of great help to investigate structure discontinuities, as for example the depression in $\Gamma_1$ observed in the second helium ionization zone, that can be used to determine the envelope helium abundance of low-mass main-sequence stars (\cite{2004MNRAS.350..277B}).

Since the method makes it possible to measure large separations depending on the parity of the degree, we can obtain odd and even values of the second differences, respectively defined as:
\begin{equation}
\seconde\ind{even/odd} = \dnu\ind{even/odd} (n+1) - \dnu\ind{even/odd} (n)
\label{seconde}
\end{equation}
From the results presented in Fig.\,\ref{fig_autodeltanuridgeHD49933}, we derive these even and odd second differences for HD\,49933 (Fig.\,\ref{fig_autodeltanu2_HD49933}). They show a very close agreement with the values derived from the fitted individual eigenfrequencies.
Fig. \ref{fig_autodeltanu2_52265} presents the variation of $\seconde$ for HD\,52265, another star observed by CoRoT (\cite{Ballot2010 }), with again a close agreement. In both cases, we remark a phase shift between the modulation of $\seconde\ind{even}$ and $\seconde\ind{odd}$. This reinforces the interest to measure the even and odd second differences according to Eq. \ref{seconde}.

\section{Conclusion\label{conclusion}}

We have shown that the envelope autocorrelation function presented by \cite{2009A&A...508..877M} and used with an ad-hoc filter is able to give the measure of the large separation of the even and odd ridges. The method is self-consistent: the EACF first provides the mean value of the large separation, then the function $\deltanunu$ derived with a narrower filter, and finally the functions $\dnup$ and $\dnui$ derived with a comb filter.
From these values, we can derive the second differences. Error bars provided by the EACF are of the same order as error bars of the mode fitting.

This method provides an alternative to mode fitting. It does not give the precise eigenfrequencies, amplitudes and lifetimes of a solar-like oscillation pattern. However, with the variation $\deltanunu$ of the large separation, with $\dnup$ and $\dnui$, and with accurate proxies of the eigenfrequencies, it gives access to a large part of the asteroseismic investigation. Its advantage, compared to mode fitting, consists in its rapidity.

\acknowledgements
I thank \'Eric Michel for the fruitful discussions we have had, that motivated this work, and Rafael Garc{\'\i}a for his useful comments on the paper.

This work is based on observations with CoRoT and \Kepler.
The CoRoT space mission was developed and is operated by the CNES, with participation of the Science Programs of ESA, ESA's RSSD, Austria, Belgium, Brazil, Germany and Spain. Funding for \Kepler\ is provided by NASA's Science Mission Directorate.


\end{document}